\documentclass{aa}  
\usepackage{graphicx,natbib}
\usepackage{txfonts}
\bibpunct{(}{)}{;}{a}{}{,} 

\newcommand{\kms}{\mbox{~km~s$^{-1}$}}

\def\dpi#1{$^2\Pi_{{#1}/2}$}

\begin{document}
  \title{Discrete Source Survey of 6\,GHz OH emission from PNe \& pPNe
  and first 6\,GHz images of K~3--35.}


   \author{J.-F. Desmurs\inst{1}
          \and
          A. Baudry\inst{2}
          \and
          P. Sivagnanam\inst{3}
          \and
          C. Henkel\inst{4}
          \and
          A.M.S. Richards \inst{5}
          \and
	  I. Bains \inst{6}
         }

   \offprints{J.-F. Desmurs}

   \institute{Observatorio Astron\'omico Nacional, C/Alfonso XII, 3,
              E-28014 Madrid, Spain, 
              \email{jf.desmurs@oan.es}
            \and  
              LAB - Observatoire Aquitain des Sciences de l'Univers, BP
              89, 33271 Floirac, France.
            \and
              GEPI, Observatoire de Paris, France.
            \and
              MPIfR, Bonn, Germany.
            \and
              Jodrell Bank Centre for Astrophysics, School of Physics
              and Astronomy, University of Manchester, UK.
	    \and
	      Swinburne University of Technology, Victoria, Australia.
             }

   \date{Received: ; Accepted: }

 
  \abstract 
{}
{The aim of this study is to investigate the physical properties of molecular
 envelopes of planetary nebulae in their earliest stages of evolution.}
{Using the 100m telescope at Effelsberg, we have undertaken a high
sensitivity discrete source survey for the first excited state of OH
maser emission ($J=5/2$, \dpi3 at 6\,GHz) in the direction of
planetary and proto-planetary nebulae exhibiting 18\,cm OH emission
(main and/or satellite lines),
and we further validate our detections using the Nan\c{c}ay radio
telescope at 1.6--1.7\,GHz and MERLIN interferometer at 1.6--1.7 and 6\,GHz.}
{Two sources have been detected at 6035\,MHz (5\,cm), both of them
are young (or very young) planetary nebulae. The first one is a
confirmation of the detection of a weak 6035\,MHz line in Vy~2--2.  The
second one is a new detection, in K~3--35, which was already known to be
an exceptional late type star because it exhibits 1720\,MHz OH
emission. The detection of 6035\,MHz OH maser emission is confirmed by
subsequent observations made with the MERLIN interferometer. These
lines are very rarely found in evolved stars. The 1612\,MHz masers
surround but are offset from the 1720 and 6035\,MHz masers which in turn
lie close to a compact 22\,GHz continuum source embedded in the optical
nebula.}
{}


   \maketitle

%

\section{Introduction}
\label{intro}

Late-type stellar objects on the post-asymptotic giant branch (AGB) are
expected to evolve through a phase, where the inner part of the stellar
envelope is becoming ionized while the outer regions are still neutral.
If the envelope is oxygen rich, such objects could exhibit OH emission
from the neutral region.  Planetary nebulae (PNe) evolve from the
envelopes of AGB stars, through the very short ($\sim$ 1000 yr)
transition phase of proto-planetary nebulae (pPNe). During this phase,
the nebular morphology and kinematics are dramatically altered: the
initially spherical, slowly expanding AGB envelope becomes a PN with,
usually, axial symmetry and high axial velocities.  More
generally, such objects may be classified as post-AGB stars. Until
recently, very few pPNe were known, Vy~2--2 being one of them.  The
ground-state hyperfine transitions of OH have been observed towards
hundreds of late-type stars and some planetary nebulae. Theoretical
studies of possible pumping mechanisms of the 18\,cm OH lines
successfully explain both the strong 1612\,MHz satellite line
\citep[see e.g.][]{elitzur76}, and, more recently, the 1665 and
1667\,MHz main lines \citep[see
e.g.][]{collison94,pavlakis96a,pavlakis96b,cragg02}.  Nevertheless,
theory is still uncertain because calculations of the OH populations in
an expanding shell are a complex task with several competing processes
which include: FIR (far infra red) pumping via the \dpi1 and \dpi3
ladders, NIR (near infra red) pumping from the radiation of the
underlying star and circumstellar dust, line overlaps of FIR lines and
collisional excitation through the $^2\Pi$ ladders.

Excited-OH hyperfine transitions have been detected up to $J=9/2$ and
$J=5/2$ in the \dpi3 and the \dpi1 ladders in the direction of several
star-forming regions \citep[for an OH level diagram, see figure~1
in][]{desmurs02}. In contrast, although the 1.6\,GHz OH lines are
easily detected towards late-type stars, the 1720\,MHz line and the
first excited state ($J=5/2$ \dpi3) are either undetected or extremely
rare and weak.

A couple of excited OH detections in evolved objects have
previously been reported, NML~Cyg at 6\,GHz ($J=5/2$, \dpi3)
\citep[][]{zuckerman72} and AU Gem at 4.7\,GHz ($J = 1/2, ^2\Pi_{1/2}$)
\citep[][]{claussen81}. However, neither detection has been confirmed
by more recent observations\citep[see for example the recent excited OH
observations in NML~Cyg by][]{sjouwerman07}.

During our previous observations of excited OH made with the Effelsberg
telescope at 6\,GHz of a sample of 65 stellar objects tracing different
stages of stellar evolution, only one source (a post-AGB star) was
detected, Vy~2--2 \citep[see figure~2 from][]
{desmurs02}. \cite{zijlstra89} conducted a discrete source survey for
OH ground state maser emission at 1.6--1.7\,GHz (18\,cm) in the direction of
several IRAS\footnote{Infrared astronomical satellite} sources. They
reported the detection of 21 possible post-AGB objects. In the
meantime, more such sources in the critical transition phase between
the AGB and the PN stage of evolution may have been identified.

We performed with Effelsberg sensitive observations of pPNe and
PNe in the 6\,GHz lines of OH.
These observations probe highly excited inner layers of the
circumstellar envelope (CSE) and they can provide important information
on the local physical conditions and on the pumping routes leading to
the OH maser phenomenon.
We followed up our new detection of 6\,GHz OH emission with sensitive
single dish observations performed at 18\,cm with the Nan\c{c}ay telescope
and with imaging at 5 and 18\,cm using MERLIN\footnote{MERLIN is the UK
radio interferometer operated by the University of Manchester on behalf
of STFC.}.

In Sect.\,2., we describe the selection criteria for the input source
sample.  The observations are reported in Sect.\,3 The results are given
and discussed in Sects.\,4 and 5., respectively. Finally the conclusions
are presented in Sect.\,6.

\section{Source sample}
In this work, we have undertaken a discrete survey of the first excited
state of OH emission ($J=5/2$, \dpi3, 6\,GHz (or 5~cm)) from post-AGB
stars with previous detections of ground-state OH maser emission. Our
sample is based on previous surveys of proto-planetary nebulae and very
young PNe \citep{zijlstra89,zijlstra01,telintel96,hu94,engels02} from
which we selected all sources north of declination --20 degrees
exhibiting 1612\,MHz emission alongside the 1665 and/or 1667 line.  In
Table \ref{table1}, we give a complete list of the 48 sources searched
by us~: 47 are post--AGB objects and one is the red Supergiant star NML
Cyg that had been suspected to show 6\,GHz emission
\citep{zuckerman72}.  With respect to our previous survey on OH stars
\citep{desmurs02}, we have repeated observations of those eight sources
that are, or are suspected to be, post-AGB sources.

\begin{table}
\begin{minipage}[t]{\columnwidth}
\caption{{\bf{Input source catalog observed at Effelsberg.}}}
\label{table1}
\setlength{\tabcolsep}{2pt}
\centering
 \begin{tabular}{|c|c|c|c|c|}
  \hline
\small
\bf{IRAS Source}&\bf{Other Name}& \multicolumn {2}{c|}{\bf{Coordinates}}&\bf{LSR
 Vel.} \\
          &               & \bf{RA(2000)}&\bf{DEC(2000)} &\bf{(km\,s$^{-1}$)} \\
  \hline
04215$+$6000&     M~4$-$18    &  04:25:50.200 & $+$60:07:11.00 & $+$38.0\\ 
04395$+$3601&               &  04:42:53.600 & $+$36:06:53.60 &  00.0\\ 
05251$-$1244&     IC~418    &  05:27:28.204 & $-$12:41:50.26 & $+$00.0\\ 
06176$-$1036&     AFGL~915  &  06:19:58.216 & $-$10:38:14.69 & $+$01.0\\ 
07396$-$1805&    NGC~2440   &  07:41:55.400 & $-$18:12:33.00 & $+$42.0\\ 
07399$-$1435& OH~231.8$+$4.2  &  07:42:16.737 & $-$14:42:14.04 & $+$35.0\\
09371$+$1212&               &  09:39:53.600 & $+$11:58:54.00 &  00.0\\
17423$-$1755&               &  17:45:14.200 & $-$17:56:47.00 & $+$50.0\\
17433$-$1750&               &  17:46:15.900 & $-$17:51:49.00 &$+$120.0\\
18091$-$1815&               &  18:12:03.400 & $-$18:14:25.00 & $+$44.0\\
18105$-$1935&               &  18:13:32.200 & $-$19:35:03.00 & $+$15.0\\ 
18135$-$1456&   OH~15.7$+$0.8 &  18:16:25.400 & $-$14:55:05.00 & $-$00.6\\
OH~12.8$-$.9&               &  18:16:49.231 & $-$18:15:01.80 & $+$61.0\\
18246$-$1032&               &  18:27:24.000 & $-$10:30:24.00 & $+$27.0\\
18276$-$1431&   OH~17.7$-$2.0 &  18:30:30.690 & $-$14:28:57.00 & $+$60.0\\
OH~24.0$-$.2&               &  18:35:43.100 & $-$08:01:33.00 &$+$108.0\\
18348$-$0526&               &  18:37:32.520 & $-$05:23:59.40 &  00.0\\
18349$+$1023&               &  18:37:19.570 & $+$10:25:33.10 & $-$21.5\\
18376$-$0846&               &  18:40:24.120 & $-$08:43:57.80 &$+$114.0\\
18491$-$0207&               &  18:51:46.900 & $-$02:04:09.00 & $+$75.0\\
18585$+$0900&               &  19:00:53.700 & $+$09:05:01.00 & $+$64.0\\
18596$+$0315&   OH~37.1$-$0.8 &  19:02:06.280 & $+$03:20:16.30 & $+$88.5\\
OH~42.3$-$.1&               &  19:09:08.200 & $+$08:16:41.00 & $+$59.0\\
19052$+$1431&               &  19:07:33.900 & $+$14:36:45.00 & $+$27.0\\
19114$+$0002&               &  19:13:58.609 & $+$00:07:31.93 &$+$100.0\\
19200$+$1035&               &  19:22:26.800 & $+$10:41:24.00 & $+$40.0\\
19204$+$0124&               &  19:22:56.980 & $+$01:30:46.90 &$+$102.0\\
OH~53.6$-$.2&               &  19:31:22.500 & $+$18:13:20.00 & $+$11.0\\
19213$+$1424&               &  19:29:37.336 & $+$22:27:16.54 & $+$40.0\\
19219$+$0947& VY~2--2        &  19:24:22.141 & $+$09:53:55.84 & $-$62.0\\
19244$+$1115& IRC$+$10420     &  19:26:48.030 & $+$11:21:16.70 & $+$78.0\\
19255$+$2123& K~3--35       &  19:27:44.000 & $+$21:30:05.00 & $+$09.0\\
19296$+$2227&               &  19:31:45.590 & $+$22:33:42.80 & $+$40.0\\
19343$+$2926& M1$-$92         &  19:36:16.767 & $+$29:32:15.80 & $-$09.0\\
19352$+$2030& OH~56.4$-$.3    &  19:37:24.000 & $+$20:36:57.80 & $+$05.0\\
19437$+$2410&               &  19:45:52.200 & $+$24:17:41.00 & $+$02.0\\
19467$+$2213&               &  19:48:51.900 & $+$22:21:15.00 & $-$44.5\\
19500$-$1709&               &  19:52:52.701 & $-$17:01:50.30 & $-$25.0\\
19566$+$3423&               &  19:58:32.280 & $+$34:31:33.70 & $-$45.0\\
20178$+$1634&               &  20:20:08.800 & $+$16:43:52.00 & $-$66.0\\
20275$+$4001&               &  20:29:24.900 & $+$40:11:21.00 & $+$00.3\\
20406$+$2953&               &  20:42:45.900 & $+$30:04:05.00 & $+$15.0\\
NML~Cyg   &               &  20:46:25.457 & $+$40:06:59.60 & $-$18.0\\
CRL~2688  &               &  21:02:18.803 & $+$36:41:38.00 & $-$40.0\\ 
21306$+$4422&    IC~511 7   &  21:32:31.000 & $+$44:35:48.00 & $-$10.0\\
22036$+$5306&               &  22:05:30.280 & $+$53:21:33.00 & $-$45.0\\
23416$+$6130&               &  23:44:03.282 & $+$61:47:22.18 & $+$38.5\\
23541$+$7031&               &  23:56:36.380 & $+$70:48:17.90 & $-$30.0\\
  \hline
  \end{tabular}
\end{minipage}
\end{table}

\section{Observations} 
\subsection {The Effelsberg survey}

With the 100m MPIfR radio-telescope at Effelsberg (Bonn, Germany), we
have undertaken sensitive 6\,GHz OH observations in May and June 2003.
The half-power beam-width of the telescope at 6\,GHz is 130''.
We used a cooled HEMT dual-channel receiver connected to only one sense
of polarization, left circular polarization (LCP). The system
temperature was $\approx$60 K ($T_{\rm mb}$) including ground pick up
and sky noise. We used a frequency switching observing mode with a
frequency throw of 244.14\,kHz. The auto-correlator AK90 was split into
four bands of 10\,MHz allowing us to simultaneously observe the two
main lines (at 6030.747 and 6035.092\,MHz) and the two satellite lines
(at 6016.746 and 6049.084\,MHz) of the OH $J=5/2$ state. There were
2048 channels per band giving a channel separation of 4.9\,kHz and thus
an effective channel separation of 0.24\kms. Proper functioning of the
system was checked by observations of the strong 6\,GHz OH emission from
the two well known compact H{\sc II} regions, W3(OH) and ON1.

Calibration of the data followed the procedure used in the 6 GHz survey
of star-forming regions made by \cite {baudry97}. OH spectra were
calibrated by observations of NGC 7027 \citep{ott94}. The signal from
the noise diode was calibrated in Jy assuming a flux density of 5.9\,Jy
for NGC\,7027. We estimate that the flux density scale uncertainty is
within 10\%.  For possible 6\,GHz radio interference, we proceeded as in
\cite {baudry97} and we simply discarded all corrupted scans. We
reached an average 3$\sigma$ noise value of about $\sim$30 mJy for each
4.9 kHz channel.

\subsection {Nan\c{c}ay observations}

We performed observations of the ground-state 18\,cm OH transitions in
K~3--35 just a couple of months after the 6\,GHz survey, in July
2003. We used the Nan\c{c}ay radio-telescope\footnote {The station de
radioastronomie de Nan\c{c}ay, France, is operated in cooperation with
the the Observatoire de Paris and the CNRS/INSU. Additional support is
provided by the R\'egion Centre.} with its FORT upgrade \footnote
{Acronym for optimized focus for the radio telescope, see
{http://www.obs-nancay.fr/nrt/a\_index.htm} for detailed
information.}. 
The half-power beam-width at 1.6--1.7 GHz is 3.5~arcmin (EW) $\times$
19~arcmin (NS) (at zero declination).  The 8192 channel digital
auto-correlator was split into 8 banks of 1024 channels each, allowing
simultaneous observations of the four ground-state transitions at
1612.231, 1665.402, 1667.359 and 1720.530\,MHz in two complementary
polarizations. Because Nan\c{c}ay is a transit instrument, the tracking
time was limited to 1 hour providing typical
3${\sigma }$ sensitivities of $\sim$0.10 Jy per channel.

We performed three runs, alternatively monitoring circular and linear
polarization. We used two frequency-switching modes, favoring either
maximum sensitivity, or double spectral coverage, for a similar
velocity resolution of 0.07\kms. Unfortunately, the data were severely
affected by radio frequency interference (RFI). The short time sampling
used in the system allowed us to selectively remove most of the
affected data, yielding reliable spectra.  However, data editing
resulted in less accurate flux and polarization calibrations. The
absorption source W12 which we observed as a calibrator was also
randomly affected. We estimate the uncertainty of the unpolarized flux
densities to be about 20\% after data editing. 

\begin{figure*}
  \centering 
  \includegraphics[width=2.2in, angle=-90]{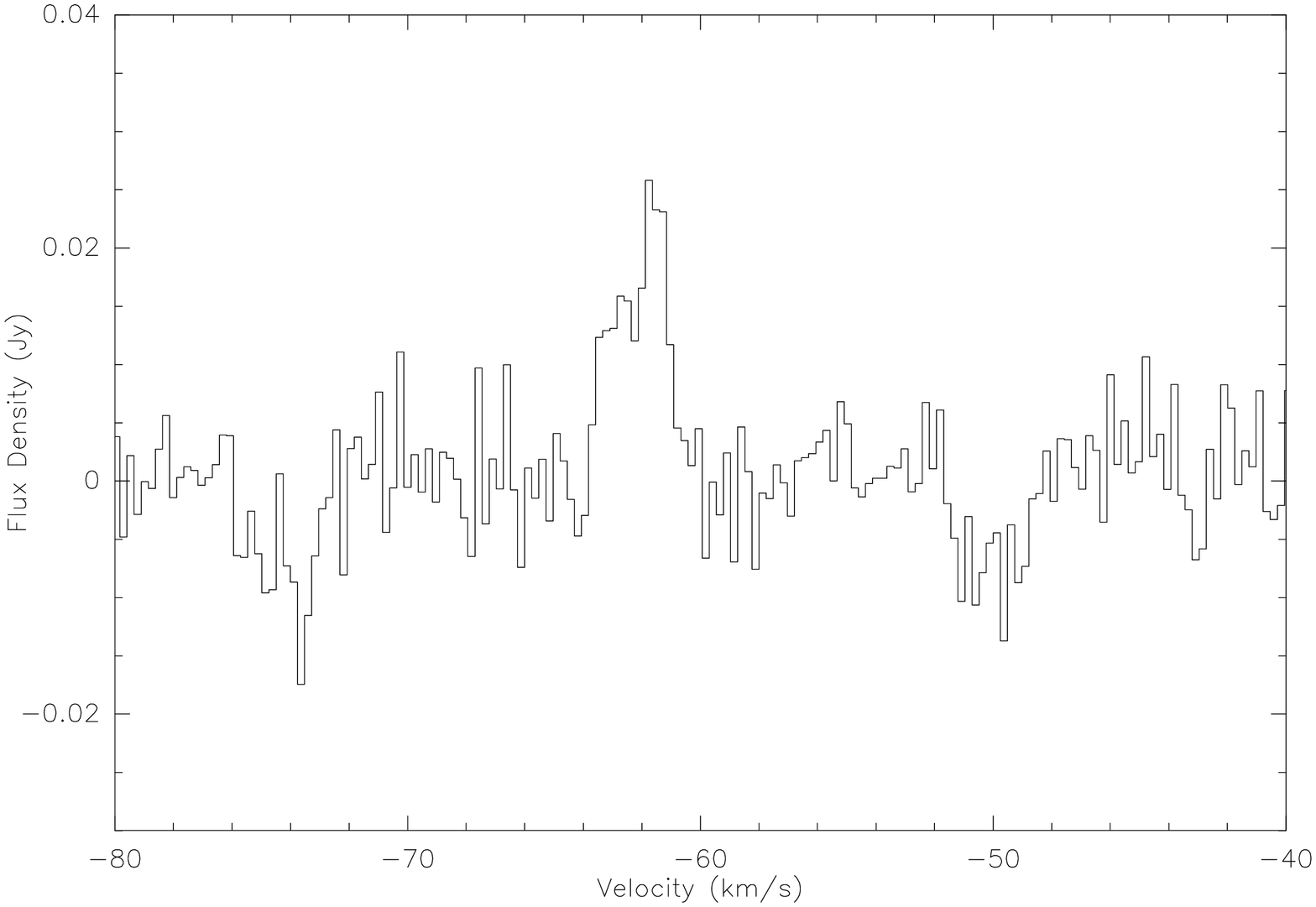}
  \includegraphics[width=2.2in, angle=-90]{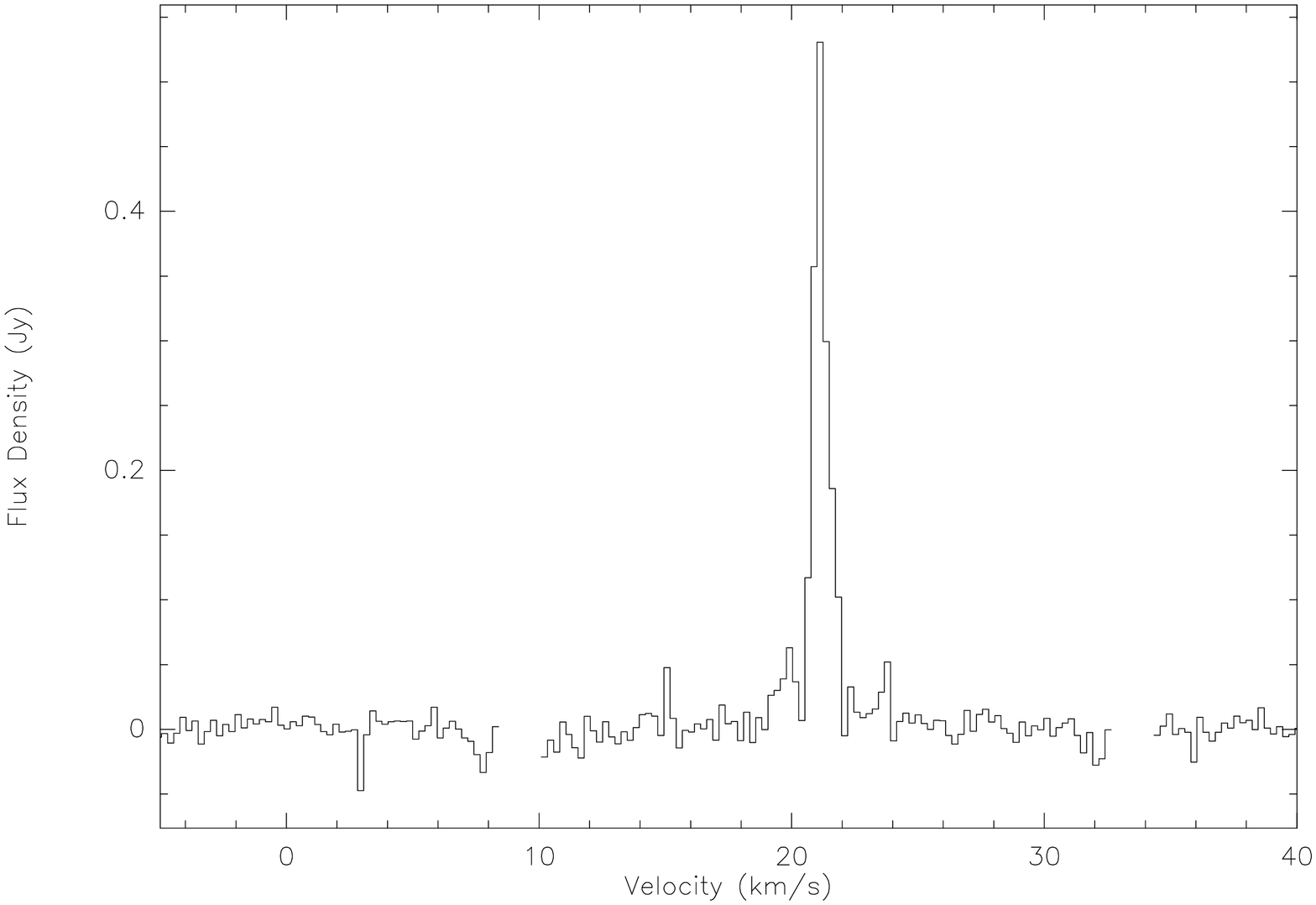}
    \caption{6035\,MHz spectra (in Jy) from Vy~2--2 (Left) and K~3--35 (Right) 
    obtained at Effelsberg.}
    \label{fig:eff6035}
\end{figure*}

\begin{figure*}
  \centering
  \includegraphics[width=3in, angle=0]{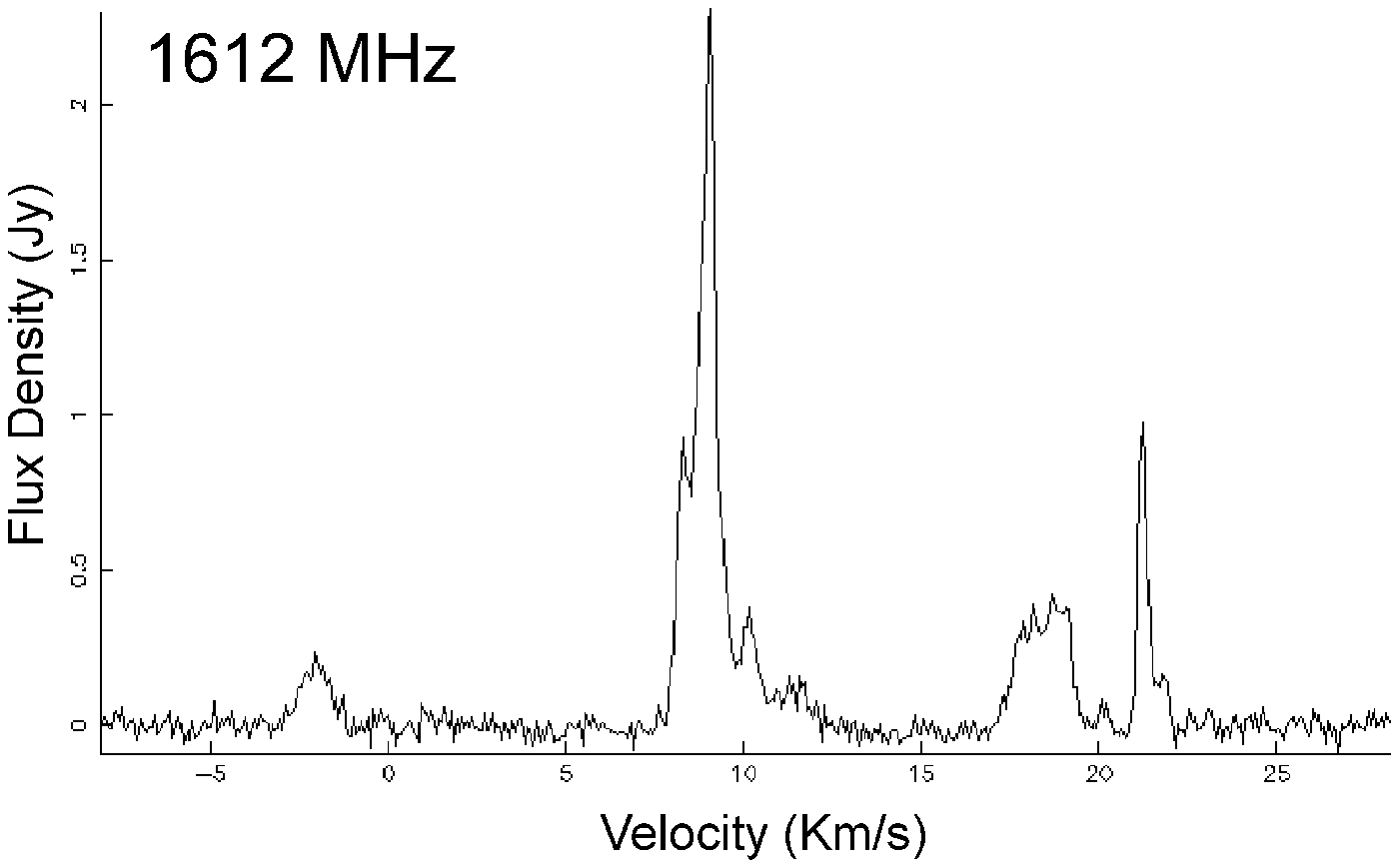}
  \includegraphics[width=3in, angle=0]{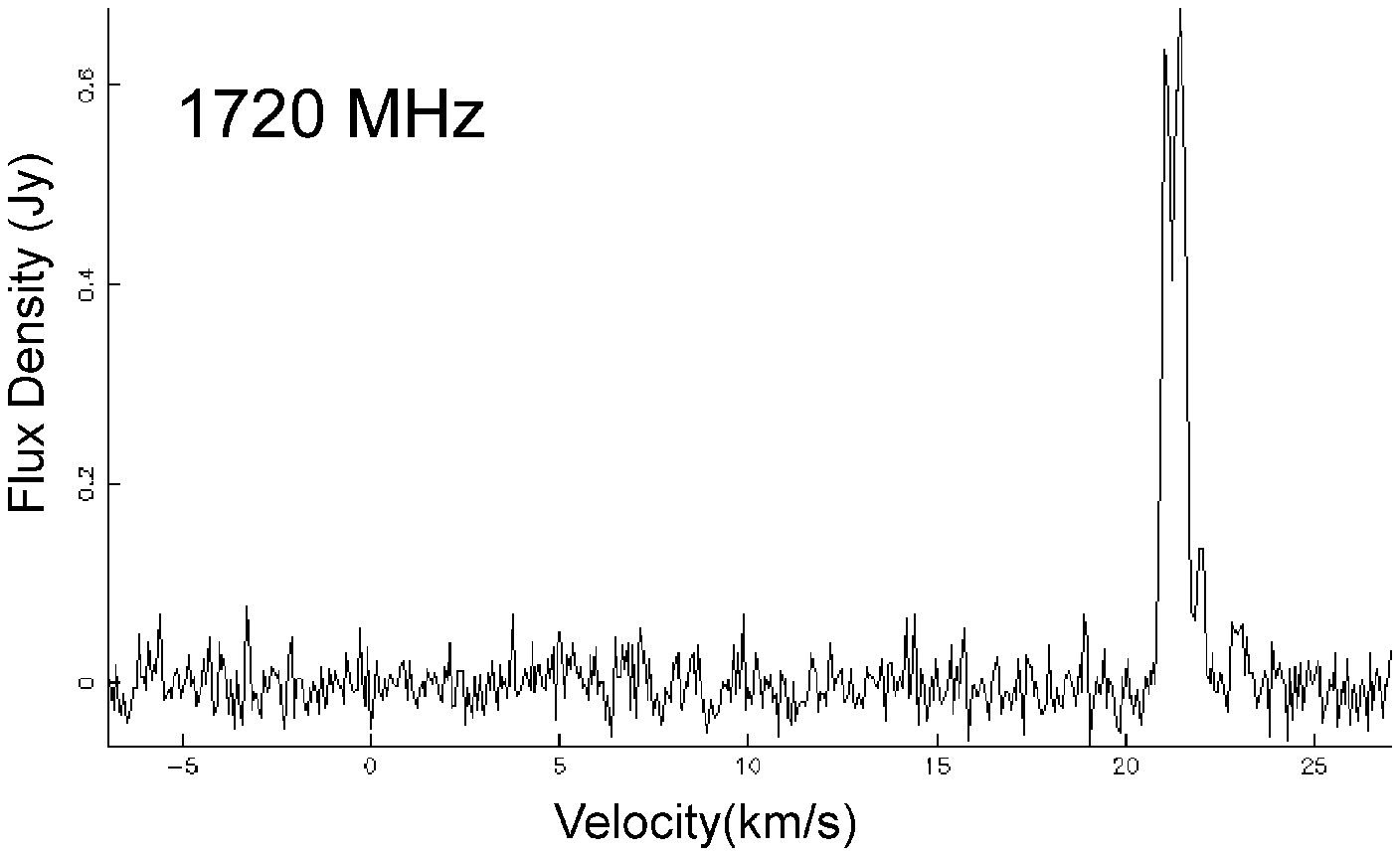}
  \includegraphics[width=3in, angle=0]{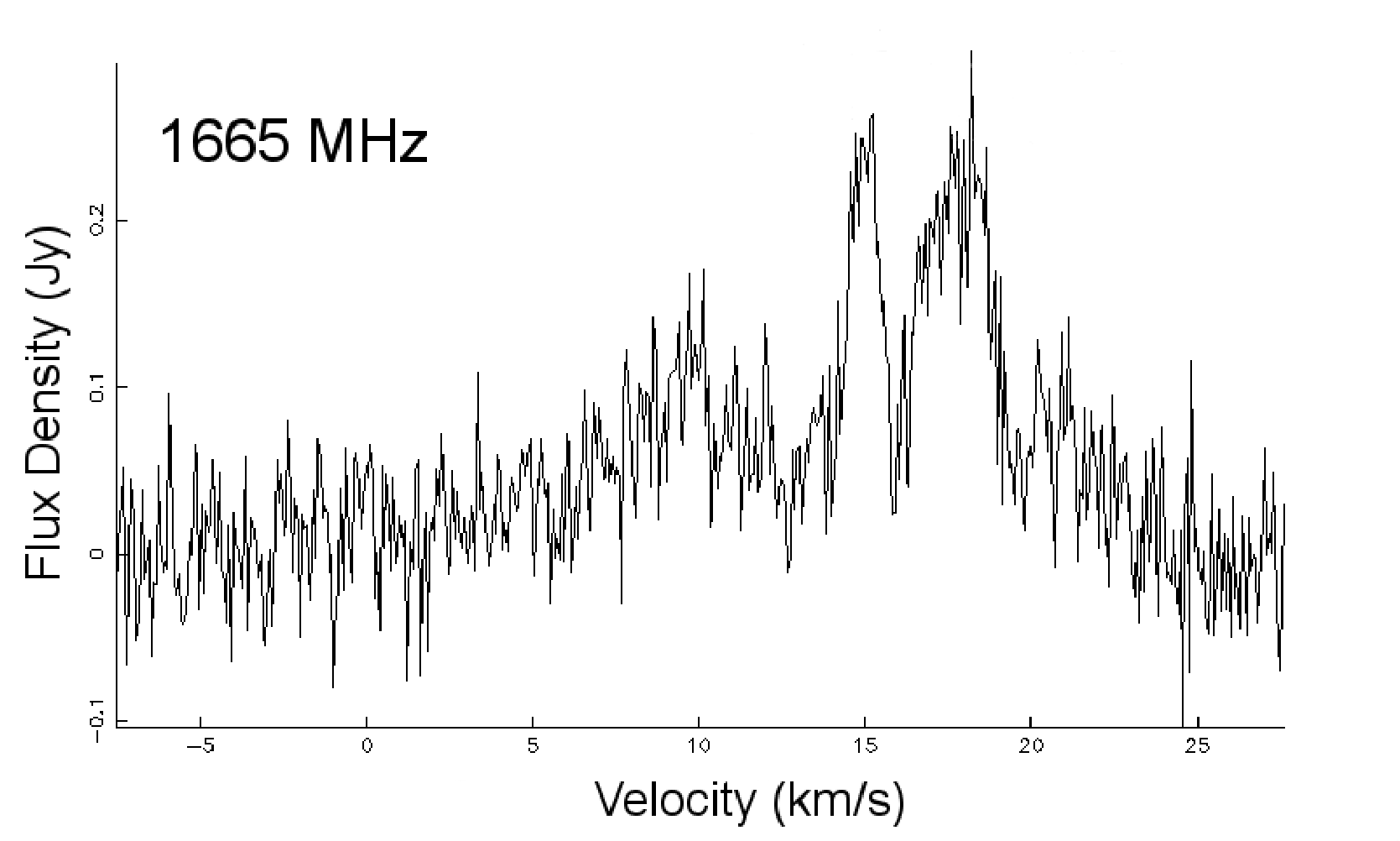}
  \includegraphics[width=3in, angle=0]{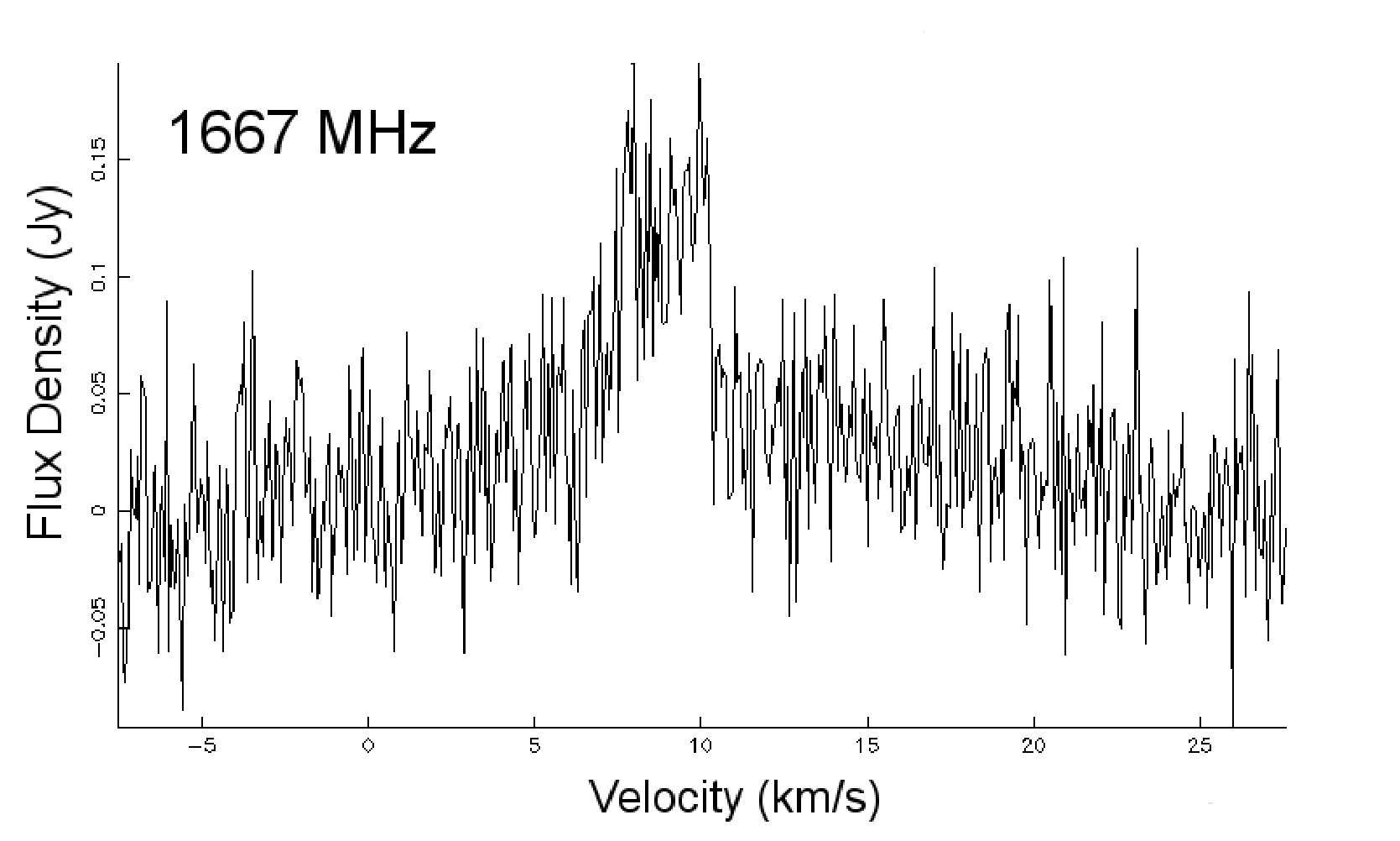}
    \caption{1612, 1720, 1665 and 1667\,MHz spectra (in Jy) from
    K~3--35 obtained at Nan\c{c}ay (sum of Left and Right polarization)
    the 1st of July 2003 less than two months after the 6\,GHz
    detection. Note that the broad pedestals in the spectra of the main
    lines (1665 and 1667\,MHz) might be an artifact from the cleaning
    process used against strong interferences (see text).}
    \label{fig:nancay1612}
\end{figure*}

\subsection {MERLIN observations}

\begin{figure*}
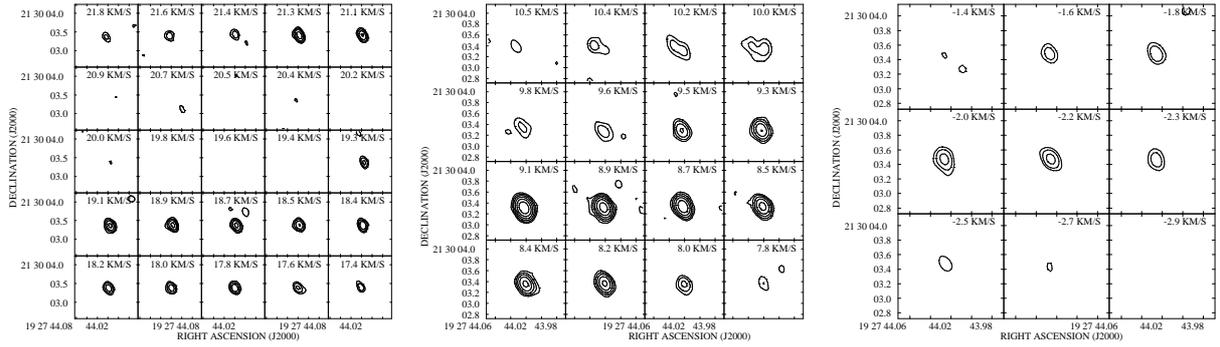

  \centering
  \includegraphics[width=2.1in, angle=0]{13387fg3a.ps}
  \includegraphics[width=2.1in, angle=0]{13387fg3b.ps}
  \includegraphics[width=2.1in, angle=0]{13387fg3c.ps}
    \caption{OH maser channels map of K~3--35 obtained with MERLIN, for
    the 1612\,MHz line covering different velocity features (with a
    clean restoring beam FWHM 226$\times $150\,mas). The peak flux is
    2.05~Jy~beam$^{-1}$ and the contour levels are
    35x(1,2,4,8,16,32,64)~mJy~beam$^{-1}$ (r.m.s per channel image
    3$\sigma \sim$ 35\,mJy/beam).}
    \label{fig:merlin1612}
\end{figure*}
\begin{figure*}
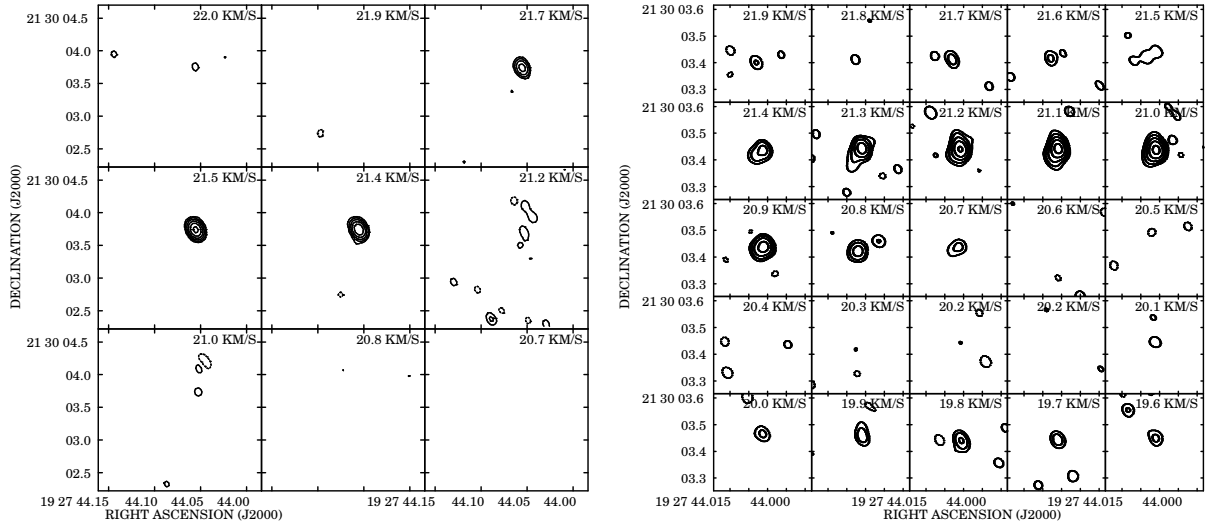

  \centering
  \includegraphics[width=2.8in, angle=-90]{13387fg4a.ps}
  \includegraphics[width=2.8in, angle=-90]{13387fg4b.ps}
    \caption{OH maser channel map of K~3--35 obtained with MERLIN, for
    the 1720\,MHz (left, clean restoring beam FWHM 177$\times $
    130\,mas) and 6035\,MHz (right, clean restoring beam FWHM 51$\times
    $ 38\,mas) transition. The peak flux densities are
    162~mJy~beam$^{-1}$(r.m.s per channel image
    3$\sigma \sim$ 15\,mJy/beam) and 781~mJy~beam$^{-1}$(r.m.s per
    channel image 3$\sigma \sim$ 90\,mJy/beam), respectively, and the
    contour levels are 4 and 35 $\times$(1,2,4,8,16,32,64)
    mJy~beam$^{-1}$. }
    \label{fig:merlin1720}
\end{figure*}


\begin{table}
\begin{minipage}[t]{\columnwidth}
\caption{Absolute positions of the strongest OH maser features
observed at 1.6--1.7 and 6\,GHz with MERLIN}
\label{table2}      
\centering          
\renewcommand{\footnoterule}{}  
\begin{tabular}{c | c | c | c | c }     
\hline 
Frequency  & R.A.   & Dec     & Pos. &LSR Vel.\\
  (MHz)    &(J2000) & (J2000) & Error &(km~s$^{-1}$)\\
           & H:m:s  & $^\circ$:$'$:$''$&(mas)&\\
\hline       
6035 & 19:27:44.0213 & $+$21:30:03.415 &$\pm$3.2& 21.1 \\
1720 & 19:27:44.0209 & $+$21:30:03.395 &$\pm$13& 21.5 \\
1612 & 19:27:44.0229 & $+$21:30:03.309 &$\pm$5.6& 9.1 \\
\hline                    
\end{tabular}
\end{minipage}
\end{table}

\begin{table}
\begin{minipage}[t]{\columnwidth}
\caption{{Relative positions, flux densities and velocities of
maser features detected with MERLIN in K~3--35. Positions are
relative to the phase center coordinates $\alpha_{\rm 2000}$ = 19$^{\rm
h}$ 27$^{\rm m}$ 44$\fs$023, $\delta_{\rm 2000}$ = $+$21$\degr$
30$\arcmin$ 03$\farcs$44. Typical relative uncertainties are given at the end
of Section 3.3.}}
\label{table3}
\setlength{\tabcolsep}{2pt}
\centering
 \begin{tabular}{|c|c|c|c|c|c|}
  \hline
\small
\bf{Component}&\multicolumn{2}{c|}{\bf{Position Offset}}&\bf{flux}&\bf{LSR Vel.}&\bf{Width} \\
           &         \bf{RA(mas)}        &\bf{DEC(mas)}     &\bf{mJy}&\bf{(km~s$^{-1}$)}&\bf{(km~s$^{-1}$)} \\
  \hline
\multicolumn {6}{|c|} {1612\,MHz} \\
  \hline
1 & $-$70.0 & $-$10.0 &  484 &  21.2& 0.25\\
2 & $-$87.0 & $-$49.0 &  395 &  18.7& 0.5\\
3 &  $-$4.0 &$-$113.0 & 2152 &   9.1& 0.5\\
4 & $+$47.0 & $+$33.0 &  212 &$-$2.0& 0.3\\
  \hline
\multicolumn {6}{|c|} {1720\,MHz} \\
  \hline
5 & $-$29.0 & $-$45.0 &  163 & 21.5 & 0.2\\
   \hline
\multicolumn {6}{|c|} {6035\,MHz} \\
  \hline
6 & $+$5.0 & $-$52.0 &  105 & 21.7& 0.2\\
7 & $-$24.0 & $-$25.0 &  950 & 21.1& 0.4\\
8 & $-$29.0 & $-$26.0 &  184 & 19.8& 0.3\\
  \hline
  \end{tabular}
\end{minipage}
\end{table}

 Taking advantage of the new capabilities of MERLIN, we observed
K~3--35 in dual circular polarization in the ground state (\dpi3,
$J=3/2$, 1.6--1.7\,GHz (or 18\,cm)) and the first excited level (\dpi3,
$J=5/2$).  All 6 antennas of MERLIN were available and all line
observations were adjusted to a central $V_{\mathrm {LSR}}$ of
12\kms. The 1612.231, 1665.402, 1667.359 and 1720.530\,MHz lines were
observed in April and May 2005, followed by 6035.092 and 6030.747\,MHz
line observations in December 2005.  The 6\,GHz observations used a
bandwidth of 0.5\,MHz giving a useful velocity coverage of 22\kms~ at a
resolution of 0.1\kms.  The 1.6--1.7\,GHz observations used 0.25\,MHz,
providing a velocity coverage of 42\kms~ at a resolution of
0.18\kms. The half-power beam-widths at 1.6, 1.7 and 6 GHz were
226$\times$150\,mas, 177$\times$130\,mas and 51$\times$38\,mas,
respectively. The 3$\sigma$ sensitivity of an individual channel was a
few tens of mJy (see Figures \ref {fig:merlin1612}, \ref
{fig:merlin1720} for precise values.)

The phase reference sources used at 6 and 1.6--1.7\,GHz were B1923+210
and B1932+204A respectively.  3C84 was used to calibrate the bandpass.
Data reduction and maser component fitting followed the procedures
described in \cite {diamond03b}, and e.g. \cite{Green07}.

Maser components with close positions in successive channels are
grouped in features, given in Table \ref {table2}, \ref{table3}.  The
relative position accuracy (derived from the beam size divided by the
signal to noise ratio) is a few milli-arcsec for the 6\,GHz features
and the brightest 1.6--1.7\,GHz features, up to 12 mas for the
faintest.  The uncertainties in comparing MERLIN observations in
different frequency bands are better than 20\,mas, dominated by errors
in transferring phase solutions from the calibration source to the
maser target \citep[see][and references therein for a fuller
description]{Green07}.

\section {Results}

\subsection {Single dish results}

From our survey of the first excited state of OH with Effelsberg, two
sources were detected.  We confirm the detection of Vy~2--2 \citep[see
also][]{desmurs02} and we detected one new source, the young PN
K~3--35, which exhibits a single feature around $V_{\rm LSR}$ = 21\kms~
in the 6035\,MHz line; the line width at half intensity is around
0.8\kms.  This OH line is the strongest ever detected at 6\,GHz toward a
late type star.  Figure \ref {fig:eff6035} shows the single dish
spectra for our two detections. The left panel shows the spectrum for
Vy~2--2. This spectrum corresponds to 2h20~min integration on-source
(5\,mJy noise level, channel width 0.24\,km\,s$^{-1}$).  The right
panel, shows our new detection in K~3--35. This spectrum was obtained
after 50 minutes on-source integration (14\,mJy noise level, channel
width 0.24\,km\,s$^{-1}$).
We did not detect the other main line at 6031 MHz, nor the satellite
lines at 6017 and 6049 MHz, in Vy~2--2, K~3--35 or any other
object\footnote{It is interesting to note that CRL\,2688, classified as
a young pPN on the basis of its strong mm-wave emission, was also
observed in this work (see Table \ref {table1}) but was not detected in
excited OH.}.  Similar results were obtained for all other 6\,GHz
OH sources observed in ultra compact H{\sc II} regions where the
6035\,MHz emission is the dominant line \citep[see][]{baudry97}.

The LTE ratio of 6035:6031 MHz emission is 20:14.  The
non-detection of K~3--35 at 6031\,MHz (above a limit of 30\,mJy)
corresponds to a ratio $>20:1$, indicating a highly non-thermal
excitation process at 6035\,MHz.  The observed single dish linewidth
(Fig. \ref {fig:eff6035}) is similar to the thermal linewidth at
200--300 K, which could indicate that the maser is approaching
saturation or is broadened by blending of more than one velocity
component.  This and other evidence for the maser nature of this line
is discussed in Section 4.2.

In Figure \ref{fig:nancay1612}, we show the sum of the two circular
polarization (RCP and LCP) obtained with Nan\c{c}ay at 18\,cm toward
K~3--35.  Both satellite lines are showing strong narrow emission
complexes. At 1720\,MHz narrow features are observed close to the
velocity of the 6035\,MHz feature whereas at 1612\,MHz, peaks are
irregularly spread over 25\kms, also extending and even peaking at more
blue shifted velocities (see also color image in
Fig. \ref{fig:k3-35rel}). In both cases, the 18\,cm features appearing
at velocities above 20\kms~ are strongly polarized (left-hand at
1612\,MHz, right-hand at 1720\,MHz).  At velocities lower than 20\kms,
polarization remains marginal.  The main lines at 1665 and 1667 were
also detected within this velocity range with a lower signal-to-noise
ratio.

\subsection {Imaging}

The relatively strong 6035\,MHz emission from K~3--35, and its strong
1.6--1.7\,GHz maser emission (see Fig. \ref{fig:nancay1612}), prompted
us to examine whether they are genuinely associated. We therefore
performed interferometric observations of K~3--35 with MERLIN.

\begin{figure}
\begin{minipage}[t]{\columnwidth}
  \centering
  \includegraphics[width=3.2in, angle=-90]{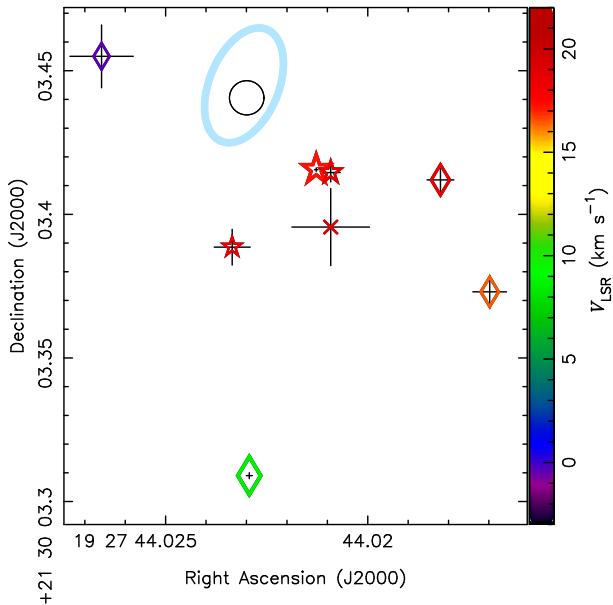}
    \caption{Relative distribution of the 6035\,MHz (star), the
  1612 (diamond) and 1720\,MHz (cross) OH emission in K~3--35. The
  maser spots have been plotted using the color velocity scale
  shown on the right side. The thin black crosses are the position
  uncertainties of each feature. The black open circle marks the
  position of the continuum emission at 1.3\,cm measured by \cite
  {miranda01}. The light blue ellipse around the continuum mark the
  position of the rotating and expanding ring of water masers found by
  \cite{uscanga08}.}
    \label{fig:k3-35rel}
\end{minipage}
\end{figure}

Our 1612\,MHz channel maps are shown in Figure \ref{fig:merlin1612} and
our 1720 and 6035\,MHz channel maps are presented in Figure
\ref{fig:merlin1720}.  The absolute positions of the strongest 6035,
1720 and 1612\,MHz features are given in Table \ref{table2}.  The
relative positions of the 6035, 1612 and 1720\,MHz maser features are
presented in Table \ref{table3} and are plotted in Figure
\ref{fig:k3-35rel}. The 6035\,MHz features, covering a velocity span of
19--22\kms, lie within about 40 mas of each other; this is an order of
magnitude larger than the relative position uncertainty achieved with
MERLIN.  The 1720\,MHz emission, at 21--22\kms, falls within 20 to 30
mas of the 6035\,MHz masers, but is spatially unresolved within the
larger position errors at the lower frequency. The 1612\,MHz masers at
similar velocities lie up to 60 mas west of the 6035\,MHz emission but
the more blue-shifted emission is extended over a much larger area of
$>150$ mas. The brightest 1612\,MHz emission, at around 9\kms, is 100
to 140\,mas south of the 6035\,MHz masers and the emission at --2\kms~
is displaced by almost as far to the NE.

\begin{figure}
\begin{minipage}[t]{\columnwidth}
  \centering
  \includegraphics[width=3.2in, angle=0]{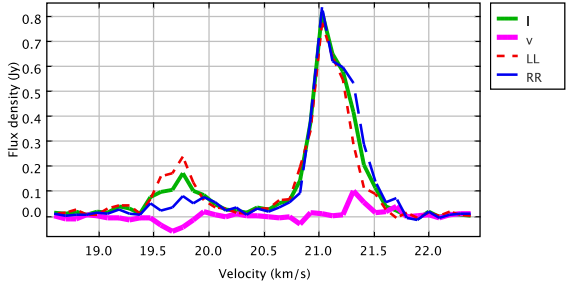}
    \caption{Flux densities averaged over the 4 pixels covering the two
brightest features at 6035\,MHz (Positions centered toward
coordinates $\alpha_{\rm 2000}$ = 19$^{\rm h}$ 27$^{\rm m}$
44$\fs$0215, $\delta_{\rm 2000}$ = $+$21$\degr$30$\arcmin$
03$\farcs$415). Stokes I and V are shown as thinner green and thicker
purple solid lines and the left- and right-hand circularly polarized
(LCP, RCP) spectra are shown as lines with short red dashes and long
blue dashes, respectively.}
    \label{fig:pol-6035}
\end{minipage}
\end{figure}

The brightest 6035\,MHz masers are less extended than the restoring beam;
this provides a lower limit to their brightness temperature of
$2\times10^7$\,K, confirming, in addition to the non LTE main line
intensity ratio, the non-thermal nature of the emission.  Maser
amplification is further suggested by the observed 6035\,MHz circular
polarization (see analysis below and Fig. \ref{fig:pol-6035}).

Observations at 6035 MHz were made in dual polarization only, so it was
not possible to carry out a full polarization calibration nor to
measure linear polarization. The ground state lines were observed in
full polarization, which will be analyzed by Bains et al. (in
prep). The 6035\,MHz data show up to 25\% circular polarization (Stokes
V as a percentage of Stokes I) at 21.3\kms, far in excess of the MERLIN
polarization leakage of 1--2\%.  Fig. \ref{fig:pol-6035} shows the flux
densities averaged over the 4 pixels covering the two brightest
features. (Because of our visibility spectra averaging only two
6035\,MHz features are visible in Fig. \ref{fig:pol-6035} whereas three
are given in Table \ref{table3}.)

Stokes I and V are shown as thinner and thicker solid lines
and the left- and right-hand circularly polarized (LCP, RCP)
spectra are shown as lines with short dashes and long dashes,
respectively. The feature at 19.8\kms~ is dominated
by LCP, possibly due to the interplay of velocity and magnetic field
gradients \citep{cook75}.
Producing the RCP and LCP maps for the brightest 6035 MHz features
observed at 21.1 km/s, we find that the two circularly polarized
components are spatially coincident.  If we interpret the frequency
shift that we were able to measure in terms of Zeeman pair (about
0.1\kms, our spectral resolution), this suggests a line of sight
magnetic field of $\lesssim 2$\,mG, although a more sophisticated
analysis is required for an accurate estimate (Bains et al. in prep).
Our upper limit is consistent with the 0.9\,mG field strength
reported by \cite{gomez09} from their 1665\,MHz maser spots found
within about 150\,AU from the 22\,GHz continuum peak (open circle in
Fig. \ref{fig:k3-35rel}).  However, we cannot exclude that our field is
stronger than that reported by \cite{gomez09}, possibly due to the
location of the 6035\,MHz emission closer to the main exciting source
than the 1665\,MHz features measured by \cite{gomez09}, or to beam
averaging leading to depolarization in the larger VLA 18\,cm beam.

\section {Discussion}

\subsection {Vy~2--2}
Vy~2--2 (G045.4--02.7) is a very young Planetary Nebula with a
kinematic age around 200\,yr \cite [see ][]{christianto98}.  Its most
recent distance estimate \citep[][]{bensby01} is 3.8~kpc \citep[see the
discussion in][]{desmurs02}. The VLA maps show a slightly elongated
continuum source at 4.885\,GHz \citep{seaquist83}. The continuum
emission originates from a compact (diameter$\sim$0.5'') and narrow
(thickness $\la$ 0.12'') shell of ionized gas.  \cite{seaquist83}
located the 1612\,MHz OH maser at the front edge of the ionized shell,
coincident with a shock front and an ionization front, placing it on
the near side of the expanding shell and thus providing an explanation
for it being blue-shifted. This suggests that the 1612\,MHz OH comes
from the remnant AGB wind, and the atypical 6035\,MHz excited state
emission is probably caused by the impact of the shock front or the
internal ionization.  The 6035\,MHz OH emission detected in this
source with a peak flux of about 20\,mJy may not be thermally excited,
but interferometric observations would be needed to prove it
definitely. This detection is consistent with that obtained by \cite
{jewell85}, and by \cite{desmurs02} and no significant variation was
observed over a time span of 20 years.

\subsection{K~3--35}

\begin{figure}
\begin{minipage}[t]{\columnwidth}
  \centering
  \includegraphics [width=\columnwidth]{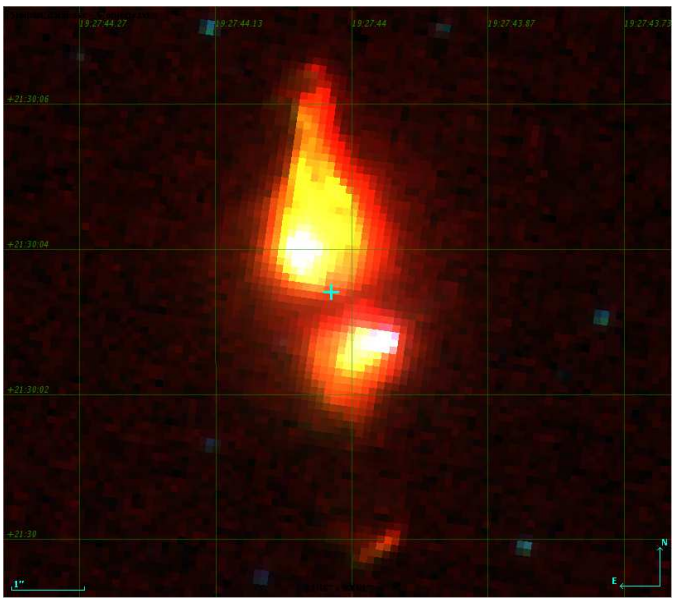}
    \caption{Composite image of K~3--35 obtained with the WFP2 camera
    from the Hubble Space Telescope. The blue cross marks the continuum
    emission position observed by \cite {miranda01}( $\alpha_{\rm
    2000}$ = 19$^{\rm h}$ 27$^{\rm m}$ 44$\fs$023, $\delta_{\rm 2000}$
    = $+$21$\degr$30$\arcmin$ 03$\farcs$44) after we cross-checked the
    HST images astrometric positions with the 2MASS catalog.}
    \label{fig:k3-35_hst}
\end{minipage}
\end{figure}

K~3--35, located at an estimated distance of 5\,kpc \citep{zhang95}, is
considered to be a very young PN \citep[][]{zijlstra89} situated in the
constellation Vulpecula and is one of the three PNe where water maser
\citep{engels85} emission at 22\,GHz has been reported \footnote {the
two others are IRAS~17347--3139, \cite{gregorio04}, and
IRAS~18061--2505, \cite{gomez08}}.  The presence of strong He{\sc II}
emission and observed intensity ratios of optical emission lines,
i.e. forbidden lines vs. hydrogen lines, are typical for PNe.  VLA
observations by \cite{miranda01} at 22\,GHz of H$_2$O emission led to
the detection of three emitting regions associated with the source, two
at the ends of the outflows (to the North and to the South) and the
third one close to the center. They also reveal the shape of a clear
bipolar structure typical of proto-Planetary Nebulae (see Figure
\ref{fig:k3-35_hst}) which allow the authors to conclude that this
object has just begun its transformation into a Planetary
Nebula. Recently, \cite{velazquez07} have modeled the morphology of
this PN by considering a precessing jet evolving in a dense AGB
circumstellar medium.
\cite{uscanga08} re-analyzed the 22\,GHz VLA data, and modeled the
kinematics of the water maser emission from the central region. They
claim the detection of a circular ring with a radius of $\sim$100~AU
with an inclination to the line of sight of 55$^o$, almost
perpendicular to the observed outflows.
In addition to the above works, emission in the 1612, 1667 and
1720\,MHz lines of OH was reported by various authors \citep[][this
work]{engels85, telintel90, gomez06}.  Recently, \cite{gomez09} mapped
with the VLA all four ground state lines of OH maser emission. They
find that the 1665 and 1720 MHz emission is spatially coincident with
the core of the nebula while the 1612 and 1667\,MHz lines originate mainly
from the extended southern lobe.  This is consistent with our 18\,cm OH
maps (see e.g. Fig. \ref{fig:k3-35rel}).

Since 6\,GHz OH lines are more commonly found in the direction of
compact H{\sc II} regions \citep {baudry97}, it is worth investigating
whether our 6\,GHz detection could be due to source contamination
within the 2 arc min beam of the Effelsberg telescope at 6\,GHz. We
first used the IPAC data base to search for nearby objects within a
radius of 2 to 3 arc min centered on IRAS~19255+2123 (K~3--35). In the
most sensitive band of the MSX survey there is one source lying well
outside the Effelsberg beam, while in the 2MASS survey there is no
bright source in the immediate vicinity of the central object. There
is, however, roughly 15 arc sec south of K~3--35, a 34~mJy radio source
listed in the 4.85\,GHz Green Bank GB6 catalog of \cite{gregory96};
\citep [see also][]{gregory91}. This faint source with about 20 arc sec
position uncertainty is most probably an extragalactic object nearly
coinciding with K~3--35.
Finally, our interferometric observations of K~3--35 show that all the
OH masers lie close to the centre of the optical nebula. We are
therefore confident that the 1612, 1720, and 6035\,MHz maser emission
is associated with K~3--35 and that the physical conditions in the
neutral gas near the nebula center resemble those found in star forming
regions thus giving rise to 6035 and 1720\,MHz line excitation.

The 1612\,MHz components detected with the Nan\c{c}ay radio telescope
between -5 and +20\kms~ do not show any 6\,GHz counterpart whereas the
emission observed around 21\kms~ is detected at 1612, 1720 and
6035\,MHz.
We note that the 6035- and 1720-\,MHz masers lie close to the 22\,GHz
continuum source detected by \cite{miranda01}(Fig. \ref{fig:k3-35rel}).
The bulk of the 1612\,MHz emission is spread between 8 to 22\kms~ and
appears offset with respect to the 6035- and 1720-\,MHz emission.  Our
1612\,MHz maser results confirm the earlier VLA observations made by
\cite{gomez09} showing that all features appear to be projected to the
southwest of the radio continuum peak with the exception of the
--2\kms~ features lying to the northeast.
The 1612, 1720 and 6035\,MHz emission is spread out across an
area of about 150\,mas diameter, corresponding to about 750\,AU if
K~3--35 is at a distance of 5~kpc.  This delineates the area in which
the densest pockets of gas are favourable to OH masing conditions.

It is difficult to conclude from our OH data alone whether the
6035\,MHz, the 1720\,MHz or the 1612\,MHz hot spots observed in K~3--35
are more closely associated with the central star, or if the star lies
somewhere in between these OH emitting regions. \cite{gomez09}
located the 1720\,MHz emission at the core of the nebula and suggested
that the same shock could excite both H$_{2}$O and 1720\,MHz OH
masers.  \cite{miranda01} measured an absolute position for the center
of the system from their VLA 1.35\,cm continuum data. Their position
matches closely our 6035\,MHz source position measurement (see Table
\ref{table2} and Fig. \ref{fig:k3-35rel}). It is thus tempting to
propose that the 6035\,MHz and 1720\,MHz emission mark the center of
the stellar system.
 We find that the 1720 and 6035\,MHz emission regions have a smaller
velocity spread and are spatially much more compact (confined within
about 30\,mas), than the 1612\,MHz emission. The different velocity
spreads observed for the different OH transitions most probably reflect
the different physical conditions required to coherently excite these
maser transitions. We note that the velocity range displayed by the
6035\,MHz emission is narrow, ranging between 19.6 to 21.9\kms, and
very close to the emission observed at 1720\,MHz. On the other hand the
1612\,MHz emission exhibits a wide velocity spread with maser emission
across nearly 25\kms.  \cite{miranda01} observed an even greater
velocity spread in their VLA data of the water maser line spots, with
features up to +36\kms~ at the southern edge of the nebula.  Assuming
the excited 6\,GHz emission indicates the velocity of the stellar
system, around 21\kms, then, the velocity spreads observed in the other
transitions and in H$_2$O from --3 to +36 km/s reflect the complex
conditions required for coherent amplification in this peculiar
expanding envelope.

\subsection{Remarks on OH pumping}
\cite{desmurs02} have briefly reviewed the excitation mechanisms for
the OH ground-state transitions in late-type stars and the possible
role of far infrared pumping. Based on the absence of 6\,GHz OH lines
from late-type stars, in contrast to several H{\sc II} regions
exhibiting 6\,GHz lines, they suggested that the absorption of 35 and
53~$\mu$m photons (see their Fig. \ref{fig:eff6035}) might be the
dominant pumping scheme.  

From the ISO data and the non detection of these IR lines in neither of
our two sources \citep[Vy~2--2 or K~3--35, see ][]{He04,He05}, we deduce
that the line to continuum ratio is, from their 3 sigma upper limits,
smaller than 0.15 and 0.19 for K~3--35 and Vy~2--2, respectively.

The exact pumping efficiency is unknown but even if it would reach 25\%
as required by most theoretical models \citep[see for
example][]{elitzur76, thai-q-tang98} or even by the model by
\cite{Gray05} who proposed an alternative route mainly using the
53~$\mu$m photons to explain the 1612\,MHz line in late type stars, we
do not expect that IR pumping is playing a dominant role.
\footnote {Assuming that the ratio between the radio and IR solid
angles is close to unity, we find that in our two sources the FIR
photons exceed the emitted 'radio photons' by a factor larger than
about 40 \citep[see discussion in][]{desmurs02}.}

The absence of detectable excited maser emission at 6\,GHz in PNe or
pPNe observed here (except for the weak emission in Vy~2--2 and
K~3--35) tends to argue in favor of a pumping scheme based on the
absorption of 35 and 53 $\mu$m photons \citep{desmurs02}. In that case
the absorption of FIR photons at 35 and 53 $\mu$m excites the OH from
the ground state to the \dpi1 levels.

However, in the two young pPNe studied here weak 6\,GHz OH emission is
present.  In the case of Vy~2--2 the ionization shell from where the
maser emission seems to originate, may present physical conditions
(shock, higher temperature and density) similar to those prevailing in
H{\sc II} regions.  We also note that satellite line emission detected
at 1720\,MHz in K~3--35 is exceptionally rare in evolved stars.
Despite the fact that this maser emission is essentially found in
supernovae remnants, at least one other case exists where 1720\,MHz
maser emission was detected in a young pPN \citep[source
OH\,0009.1--0.4, see][]{sevenster01}. For that source, the authors
argue that a post-AGB star might show such an emission just after it
evolves from the thermally pulsing phase. We further note that isolated
1720\,MHz maser sources have been observed in several star forming
regions \citep[see][]{etoka05,edris07}.  This suggests that a hybrid
pumping model applying to both some well evolved stars and H{\sc II}
regions, or even a pumping scheme similar to that required for OH maser
emission in massive star forming regions \citep[see][]{pavlakis96a,
pavlakis00}, may be successful to explain the OH properties of the two
pPNe studied here.  In addition, around +21\kms, marking the 6\,GHz
emission in K~3--35, strong, opposite circular polarizations were found
for the 1612 and 1720\,MHz components, whereas the rest of the
1612\,MHz components were almost unpolarized.  This fact could suggest
the merging of two distinct circumstellar structures with possibly
different pump mechanisms, parameter ranges, and/or distances to the
star, although any large separation between sources at similar velocity
remains puzzling. Consistently, \cite{miranda01} displayed noticeable
polarization of the 1665\,MHz masers, which they located near the
center of the system.

In fact, the maser emission observed in Vy~2--2 and K~3--35 is
atypical.  Both sources show OH emission from the ground-state of OH
and from the 6\,GHz excited state. It is hard to deduce relevant
physical parameters from existing OH excitation models. Nevertheless,
we stress that the 1612 and 1720\,MHz lines are observed simultaneously
but at different locations. From \cite {caswell99}, the ground state
satellite lines (1612 or 1720\,MHz) appear to be most often associated
with the 1665\,MHz ground state main line, but nearly never at the same
time. Theoretical works treat the 1612 and 1720\,MHz lines as
complementary with one line being commonly inverted while, in the same
volume, the other one is in absorption. In \cite{cragg02}, it is found
that 1612\,MHz appears in a zone of high density ($n_H >
10^6$\,cm$^{-1}$), for a high OH column density ($N_{OH}/\Delta{V} >
10^{11}$\,cm$^{-3} s$) and a high gas temperature T$_k >$100K. However,
we presume that such conditions do not apply well to K~3--35.  Although
they are clearly associated with K~3--35, the 1612 and 1720\,MHz lines
are well separated (see position offset in Table \ref{table3} and
Fig. \ref{fig:k3-35rel}) and, contrary to models of OH in the
interstellar medium, both lines are seen in emission thus suggesting
that we indeed observe specific OH excitation in two distinct
circumstellar structures.

\section{Summary}

We have detected an excited-state 6\,GHz OH maser in the pPNe K~3--35,
as part of an extensive survey of 47 northern (p)PNe with ground-state
OH maser emission.  The only other object detected at 6\,GHz was the
known emitter Vy~2--2. These results reveal that 6\,GHz emission is
exceptional from such objects and suggest that K~3--35 and Vy~2--2
exhibit a peculiar stage of structure and/or evolution.  Follow-up
MERLIN interferometric observations of K~3--35 show that the 6035\,MHz
emission is very compact and located close in angular separation and
velocity to the 1720\,MHz maser line, while the 1612\,MHz line appears
to be offset by $\gtrsim$0.1 arcsec. We also note that the 1720\,MHz
line is very rarely found in post-main sequence objects. Finally, we
suggest that the 1720 and 6035\,MHz line emissions are closely
associated with the center of the stellar system.

\acknowledgements {The authors would like to thank Dr. M.Wang for her
kind help during the observations with the Effelsberg telescope.  This
work makes use of EURO-VO software, tools or services. The EURO-VO has
been funded by the European Commission through contract numbers
RI031675 (DCA) and 011892 (VO-TECH) under the 6th Framework Programme
and contract number 212104 (AIDA) under the 7th Framework Programme.
We gratefully thank the anonymous referee for his useful comments.}

\bibliographystyle{aa} \bibliography{references}
\end{document}